\documentclass[%
 reprint,
superscriptaddress,
 amsmath,amssymb,
 aps,
 prl,
]{revtex4-1}

\usepackage{mathtools}          %loads amsmath as well
\usepackage{siunitx}          
\usepackage{lipsum}
\usepackage{graphicx}% Include figure files
\usepackage{dcolumn}% Align table columns on decimal point
\usepackage{bm}% bold math
\usepackage[dvipsnames]{xcolor}
\usepackage[colorlinks=true,linkcolor=Blue,urlcolor=Blue,citecolor=Blue]{hyperref}

\def\approxprop{%
  \def\p{%
    \setbox0=\vbox{\hbox{$\propto$}}%
    \ht0=0.6ex \box0 }%
  \def\s{%
    \vbox{\hbox{$\sim$}}%
  }%
  \mathrel{\raisebox{0.7ex}{%
      \mbox{$\underset{\s}{\p}$}%
    }}%
}

\begin{document}

\title{Phase-stable self-modulation of an electron-beam in a magnetic wiggler}

\author{James P. MacArthur}
\email{jmacart@slac.stanford.edu}
\affiliation{SLAC National Accelerator Laboratory, Menlo Park, California 94025, USA}
\affiliation{Stanford University, Stanford, California 94305, USA}

\author{Joe Duris}%
\affiliation{SLAC National Accelerator Laboratory, Menlo Park, California 94025, USA}

\author{Zhen Zhang}%
\affiliation{SLAC National Accelerator Laboratory, Menlo Park, California 94025, USA}

\author{Alberto Lutman}
\affiliation{SLAC National Accelerator Laboratory, Menlo Park, California 94025, USA}

\author{Alexander Zholents}
\affiliation{Argonne National Laboratory, Lemont, Illinois 60439, USA}

\author{Xinlu Xu}
\affiliation{SLAC National Accelerator Laboratory, Menlo Park, California 94025, USA}

\author{Zhirong Huang}%
\affiliation{SLAC National Accelerator Laboratory, Menlo Park, California 94025, USA}
\affiliation{Stanford University, Stanford, California 94305, USA}

\author{Agostino Marinelli}%
\email{marinelli@slac.stanford.edu}
\affiliation{SLAC National Accelerator Laboratory, Menlo Park, California 94025, USA}

\date{\today}

\begin{abstract}
Electron-beams with a sinusoidal energy modulation have the potential to emit sub-femtosecond x-ray pulses in a free-electron laser. The energy modulation can be generated by overlapping a powerful infrared laser with an electron-beam in a magnetic wiggler. Here we report on a new infrared source for this modulation, coherent radiation from the electron-beam itself. In this self-modulation process, the current spike on the tail of the electron-beam radiates coherently at the resonant wavelength of the wiggler, producing a six-period carrier-envelope-phase (CEP) stable infrared field with gigawatt power. This field creates a few MeV, phase-stable modulation in the electron-beam core. The modulated electron-beam is immediately useful for generating sub-femtosecond x-ray pulses at any machine repetition rate, and the CEP-stable infrared field may find application as an experimental pump or timing diagnostic.
\end{abstract}

\maketitle

The first generation of x-ray free-electron lasers have now operated for a decade~\cite{Emma2010,Ishikawa2012,Ko2017,Milne2017,Tschentscher2017}, supplying gigawatt x-ray beams to a variety of users~\cite{Bostedt2016}. These facilities typically lase by self-amplifying spontaneous emission (SASE), a process which produces longitudinally incoherent beams whose spectral widths lie in the range of $\Delta \omega/\omega \approx 10^{-3}-10^{-4}$~\cite{Kondratenko1980,Bonifacio1984} and whose longitudinal signatures match that of the electron-beam current at few to hundreds of femtoseconds in duration. 

There is interest from the community of x-ray laser users in pulses capable of probing phenomena with sub-femtosecond resolution~\cite{Krausz2009,Bucksbaum2007}. Successful experimental efforts~\cite{Emma2004,Huang2017,ROSENZWEIG2008,REICHE2008,Ding2009,Prat2015,Lutman2016} toward this goal have yet to break the sub-femtosecond barrier at soft x-ray energies.

Single-spike, sub-femtosecond x-ray beams may be produced by electron-beams with a nearly single-cycle energy modulation~\cite{saldin2006,zholents2005,Tanaka2013,shim2018}. These beams could be generated by overlapping a single-cycle carrier-envelope-phase (CEP) stable laser with an electron-beam in a wiggler~\cite{hemsing2014}. Suitable infrared lasers exist~\cite{fu2018}, but challenges in optical transport and laser-electron synchronization hinder progress.

In this letter we demonstrate that an electron-beam may be modulated in a six-period wiggler with no external laser present at the Linac Coherent Light Source (LCLS). Instead, coherent radiation from a current spike on the electron-beam tail creates a quasi-single-cycle energy modulation in the beam-core. The modulation exhibits sub-femtosecond stability and is a few MeV in amplitude, in agreement with a line-charge model~\cite{SALDIN1997,SALDIN1998,Wu2003}, a paraxial model developed here, and the 3D code \textsc{osiris}~\cite{Fonseca2002}. These beam characteristics are sufficient for enhanced-SASE operation at any repetition rate.

A six-cycle, CEP-stable, gigawatt infrared pulse is a byproduct of this process. The pulse is timed with sub-femtosecond precision relative to the electron-beam, and could therefore be used as timing fiducial or in pump-probe experiments. 

\begin{figure}[t]
   \centering
   \includegraphics{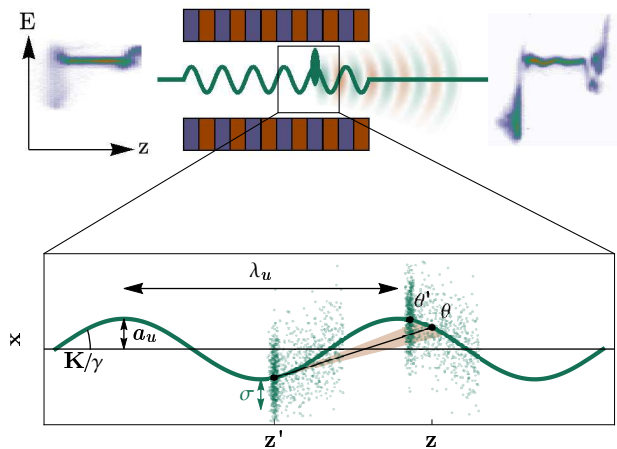}
\caption{(top) An electron-beam enters a six-period wiggler. Radiation generated in the wiggler interacts with the beam, producing a sinusoidally modulated phase space. (bottom) Inside the wiggler, a single electron-beam (green dots) of rms width $\sigma$ traverses a sinusoidal path of amplitude $a_u$ from left to right. The high current tail slice of the beam, $\theta'$, emits radiation at the longitudinal position $z'$ that reaches a core slice $\theta$ at the longitudinal position $z$.}
   \label{f1}
\end{figure}

A schematic of our experiment is shown in Fig.~\ref{f1}. A beam of electrons with relativistic factor $\gamma$ travels left to right along a sinusoidal path (green) of wavelength $\lambda_u = 2\pi /k_u$ within a six-period planar wiggler. The resonant wavelength in the wiggler is
\begin{equation}
    \lambda_1 = \frac{\lambda_u}{2\gamma^2} \left( 1+\frac{K^2}{2}\right) \approx \frac{\lambda_u K^2}{4 \gamma^2},
\end{equation}
where the planar-wiggler deflection parameter, $K$, satisfies $0\ll K \ll \gamma$. The tail of the electron bunch, a current spike shorter than the resonant wavelength in the wiggler, emits coherently at the wavelength $\lambda_1=2\pi / k_1$.  This radiation resonantly modulates the beam-core as it slips ahead of the electrons. 

When the oscillation amplitude, $a_u = K/\gamma k_u\approx \sqrt{\lambda_u \lambda_1}/\pi$, is much larger than the transverse beam-width, $2 \sigma$,
\begin{equation}
    \hat{\sigma} = 2 \sigma/a_u = \sigma \sqrt{k_1 k_u} \ll 1,
    \label{smallbeamlim}
\end{equation}
a line-charge model developed from the Li\'enard--Wiechert fields~\cite{SALDIN1997,SALDIN1998,Wu2003} adequately describes the self-modulation process. This model includes short-range space-charge-like effects and long-range radiative effects~\cite{Geloni2007}. It is relevant to our experiment, wherein $\hat{\sigma}\sim 0.4$. The energy-modulation in this model is complicated by a numerical solution to a transcendental equation, and the line-charge assumption prevents the understanding of beam-size dependence. We therefore develop a simplified 3D paraxial model that serves a complimentary role in explaining self-modulation.

We are primarily interested in the radiative modulation of the beam-core. We demonstrate in the supplemental materials that the near-axis field described in FEL theory texts~\cite{kim2017} is sufficient for calculating the radiation from the tail capable of modulating the beam-core. We also consider only the transverse radiative field under the slowly varying envelope approximation. Our analysis therefore fails to reproduce short-range effects, but does correctly predict the modulation in the beam-core. 

The paraxial equation for the near-axis, wiggle-averaged, slowly-varying field amplitude $\mathcal{E}$ resulting from an electron-beam containing $N_e$ electrons is~\cite{kim2017}
\begin{eqnarray}
    \left[ \frac{\partial }{\partial z} + k_u \frac{\partial}{\partial \theta} -  i\frac{\nabla_\bot^2}{2k_1}  \right] \mathcal{E}(\mathbf{x},\theta;z) =\nonumber \\ -\kappa_1 k_1 \sum_{j=1}^{N_e} e^{-i \theta_j(z)} \delta (\mathbf{x} - \mathbf{x}_j),
    \label{field}
\end{eqnarray}
where $z$ is the propagation distance in the lab frame,
$\kappa_1= e K \text{[JJ]}/4 \epsilon_0 \gamma$ is a coupling constant, $\text{[JJ]} = J_0(1/(2+4/K^2)) - J_1(1/(2+4/K^2))$, $\theta_j = (k_1+k_u)z - \omega_1 \bar{t}$ is the ponderomotive phase of the $j^{\text{th}}$ electron at a wiggler-averaged arrival time $\bar{t}$, and $\mathbf{x}_j = (x_j,y_j)$ is the transverse position of the $j^{\text{th}}$ electron. We adopt the notation of~\cite{kim2017} wherein the horizontally polarized electric field is $\mathbf{E} = \left(\mathcal{E}e^{i\theta} + \mathcal{E}^* e^{-i \theta}\right) \hat{x}$. Equation~\ref{field} only includes radiation at the fundamental harmonic, $\lambda_1$. In our experiment harmonic content is suppressed by a current spike larger than $\lambda_1/3$ in extent. 

A Green's function for the first-harmonic modulation from an arbitrary current distribution may be calculated from the field produced by a delta-function distribution in the longitudinal dimension, $\theta_j(z) = 0$. If the transverse distribution of the beam is a symmetric Gaussian, $\sum \delta (\mathbf{x} - \mathbf{x}_j) \propto \exp\{-(x^2+y^2)/(2\sigma^2)\},$ Equation~\ref{field} can be solved without approximation. Given the boundary condition that there is no electric field at the start of the wiggler, $\mathcal{E}(\mathbf{x},\theta;z=0)=0$, the field amplitude is
\begin{equation}
\mathcal{E}=
      \begin{cases}
         -\frac{2\pi N_e \kappa_1}{\lambda_1^2 \left(i \theta + \hat{\sigma}^2\right)} \exp \left( -\frac{1}{2}\frac{ \hat{x}^2 + \hat{y}^2}{i \theta + \hat{\sigma}^2} \right), & \text{$0<\theta<k_u z$} \\
         0, & \text{otherwise} 
  \end{cases}
  \label{solution}
\end{equation}
where $\hat{x}^2 = k_1 k_u x^2$ and $\hat{y}^2 = k_1 k_u y^2$.
Equation~\ref{solution} can be recognized as the $\text{TEM}_{00}$ Gaussian mode. This equation describes the field at angles relevant to self-modulation, but other references~\cite{hofmann2004,wiedemann2007,Geloni2005} should be consulted for a complete description of the radiation field. % Evidently a Gaussian electron-beam source creates a Gaussian radiation beam. The radiation beam is born at its waist of size $\sigma$. It executes half a Gouy phase shift as it diffracts and drifts forward relative to the electrons. Radiation created at $z=0$ has only reached a phase of $k_u z$ after a travel distance of $z$ in the wiggler -- no field exists beyond $k_u z$. 
%\iffalse
%The power of this beam is 
%\begin{equation}
%    P = \frac{2}{\mu_0 c} \int \left|\mathcal{E}\right|^2 dxdy = \frac{2\pi N_e^2 \kappa_1^2}{\mu_0 c \lambda_1^2 k_u^2 \sigma^2},
%    \label{eqpower}
%\end{equation}
%which is a few gigawatts for LCLS-like parameters. 
%\fi
This carrier-envelope-phase stable few cycle pulse is coincident with the electron-beam and powerful enough to modulate the electrons it passes through. The field-induced relative energy modulation, $\Delta \gamma / \gamma = \eta$, grows in proportion to the field~\cite{kim2017},
\begin{eqnarray}
\frac{d \eta(\mathbf{x},\theta;z)}{dz} = \chi_1 \left(\mathcal{E} e^{i \theta} + \mathcal{E}^* e^{-i \theta}\right),
\label{emode}
\end{eqnarray}
where $\chi_1 = e K \text{[JJ]}/2\gamma^2 m c^2$ is a coupling factor. Equation \ref{emode} is a wiggler-period-averaged expression, meaning non-resonant terms that vary rapidly over a wiggler period have been dropped. The expression may be integrated over $z$ and averaged over the transverse distribution to find the mean energy deviation
\begin{equation}
    \langle \eta(\theta;z) \rangle = - \eta_0 (k_u z -\theta) \frac{2 \hat{\sigma}^2 \cos \theta + \theta \sin \theta}{\theta^2 + 4 \hat{\sigma}^4},
    \label{emodfull}
\end{equation}
when $\text{$0<\theta<k_u z$}$, and $\langle \eta(\theta)\rangle=0$ otherwise. The amplitude  $\eta_0 = 2 k_1 N_e r_e\text{[JJ]}^2/\gamma$ is most compactly written in terms of the classical electron radius $r_e$. It can be several MeV under LCLS-like conditions, exceeding the slice energy-spread of the electron-beam. Our formalism is only valid in the beam-core where $\theta > 2\pi$, a region in which
\begin{equation}
\langle \eta(\theta;z) \rangle =
      \begin{cases}
         - \eta_0 (k_u z -\theta)\, \text{sinc}\, \theta, & \text{$0<\theta<k_u z$} \\
         0, & \text{otherwise} 
  \end{cases}
  \label{emod}
\end{equation}
for $\hat{\sigma}=0.4$. The disappearance of $\hat{\sigma}$ in this regime suggests Equation~\ref{emod} should be consistent with the line-charge model. We demonstrate this agreement in the supplemental materials. We also discuss the effect of a large beam-width in the supplemental materials.  %The Gouy phase shift discussed below Equation~\ref{solution} manifests in Equation~\ref{emod} as a modulation phase shift. 
%Eq. \ref{emod} confirms the intuitive explanation of the process given above: when $\hat{\sigma}^2 \ll 1$, the $\cos{\theta}$ term is negligible for distances of one wavelength of more away from the source and the modulation is dominated by the decaying term $(k_u z -\theta)\sin(\theta)/\theta$, which produces a quasi-single-cycle modulation.

Two components of Equation~\ref{emod} contribute to the decay of the energy modulation along the bunch. The dominant term is the denominator in $\text{sinc}\, \theta = \sin (\theta) / \theta$, a result of diffraction. The factor $(k_u z -\theta)$, a result of the non-infinite wiggler extent, represents a shortened interaction length for slices of the bunch far from the tail. Electrons $\theta/2\pi$ wavelengths away from the source only interact with the radiation over $(k_u z - \theta)/2\pi$ wiggler periods. These two effects conspire to produce a quasi-single-cycle energy modulation.

Self-modulation was observed experimentally using an X-band Transverse deflecting Cavity (XTCAV)~\cite{behrens2014} at LCLS. The LCLS current profile has spikes at the head and tail of the bunch that are typically suppressed with two collimators in a dispersive section~\cite{ding2016}. In this experiment we remove one collimator to maximize the peak-current in the tail.

\begin{figure}[htb]
   \centering
   \includegraphics{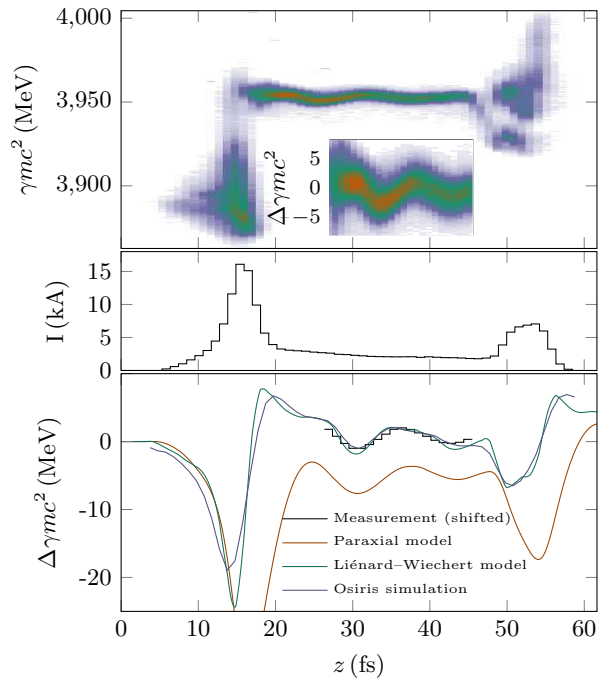}
\caption{(top) The measured transverse phase-space of a modulated electron-beam with the tail to the left. The inset shows the beam-core. (middle) A projection onto the time-axis yields the current profile. (bottom) The energy modulation predicted from the paraxial model (red) the Li\'enard--Wiechert model (green), and the \textsc{osiris} simulation (blue) match the shifted measurement data (black).}
   \label{f2}
\end{figure}

In Fig.~\ref{f2} the electron-beam phase-space from an XTCAV measurement is projected onto the longitudinal axis to reveal a high current spike on the electron-beam tail. Wiggler and beam parameters are given in Table~\ref{table}. This current profile is convolved with Equation~\ref{emodfull} to produce a modulation profile. The current is also convolved with the Green's function for a line-charge model based on the Li\'enard--Wiechert fields~\cite{SALDIN1998,Wu2003}. Evidently the paraxial model fails to reproduce the effect of the short-range wake in the current spike, but it does correctly represents the energy-modulation in the beam-core where conditions are favorable for enhanced-SASE. The downward displacement of the paraxial model is also a result of the incorrect short-range behavior of Equation~\ref{emodfull}. We note the compression factor $R_{56}=\Delta z/\eta$ of the wiggler perturbs the current profile over 6 periods, a phenomenon not accounted for with these models. This effect is important for longer wigglers~\cite{Zhang2019}.

We also performed a simulation of the self-modulation process in the beam rest-frame using the 3D particle-in-cell code \textsc{osiris}~\cite{Fonseca2002}. The simulation inputs were the current distribution in Fig.~\ref{f2}, a $\SI{2}{MeV}$ rms slice-energy-spread, a beam size of $\hat{\sigma}=0.4$, a normalized emittance of $\SI{0.4}{\micro \meter}$, and wiggler parameters given in Table~\ref{table}. The simulation time-step and grid-size were set to resolve short-range effects, $dx=dy=dz=2c \, dt=1/(16 k_1')$, with $k_1'$ representing the resonant wavenumber in the average rest-frame. The average rest-frame is boosted along the $z$-axis by a Lorentz-factor of $\gamma_z = \gamma /\sqrt{1+K^2/2} = 212.3$ relative to the lab frame. The output energy modulation is in close agreement with the line-charge model.

The measured energy modulation from XTCAV is shown in black for comparison. Due to challenges in reconstructing the exact phase-space before the wiggler, the data were shifted vertically and horizontally for comparison purposes. Only the modulation amplitude and wavelength should be inferred from these data.

\begin{table}[b]
\caption{\label{table}%
Wiggler and electron-beam parameters.
}
\begin{ruledtabular}
\begin{tabular}{lllll}
Parameter & Figs.~2, 3 & Fig.~4 & Fig.~5 \\
\colrule 
           Wiggler gap (mm)     & 8.2 & 8.2 &11.5\\ 
           Wiggler K-value\footnote{Calculated from hall-probe field-maps.}  &    51.5   & 51.5  &  43.3\\ 
           Wiggler period (cm)  &  35     & 35  & 35\\ 
           Wiggler length\footnote{Includes fringe fields, effective magnetic length is 6 periods.} (cm) &     230  & 230   & 230\\ 
           Beam energy\footnote{Reported values are set-points, actual values vary shot-to-shot.} (MeV)    &   3953      & 3782 & 3420\\ 
           Beam charge\textsuperscript{c} (pC)    &  200       & 180  & 140
\end{tabular}
\end{ruledtabular}
\label{paramtable}
\end{table}

\begin{figure}[htb]
   \centering
   \includegraphics{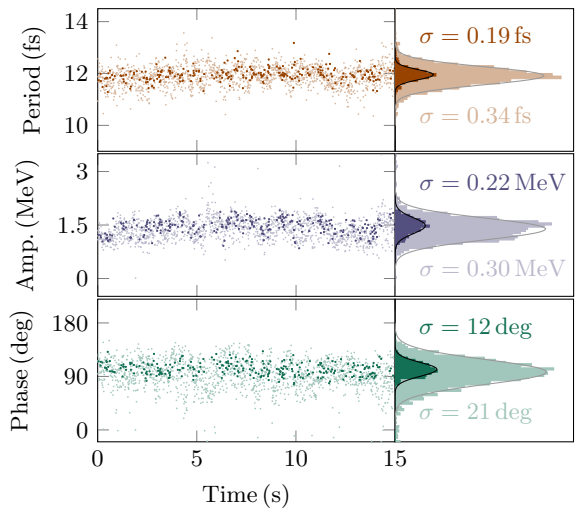}
\caption{The best fit modulation period (top), amplitude (middle), and phase (bottom) for consecutive pulses in a 15 second timeframe. The rms width of a Gaussian fit to the binned data is also reported (right), for both the entire data set (light), and a data set where the peak current in the second bunch compressor is restricted (dark).}
   \label{f3}
\end{figure}

To demonstrate the stability of self-modulation we measure the variation in the modulation period, amplitude, and phase relative to the current spike for a series of 1800 consecutive shots in Fig.~\ref{f3}. The raw rms modulation period variation is $\SI{340}{as}$. Much of this variability is due to the $\SI{1.06}{fs}$ temporal resolution of the TCAV and linac jitter that modifies the peak current on a shot-by-shot basis. After filtering by the peak-current as measured in the second bunch compressor, the rms period variability drops to $\SI{190}{as}$. Similar improvements are seen in the modulation amplitude and phase. 

The intrinsic stability of the self-modulation process makes it a reliable replacement for modulation from an external laser. 

\begin{figure}[htb]
   \centering
   \includegraphics{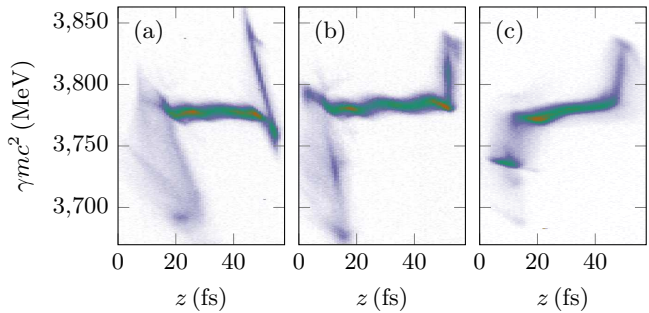}
\caption{The electron-beam phase-space as measured in the dump when the dog-leg 2 $R_{56} = \SI{-0.25}{mm}$ (a), $ \SI{-0.215}{mm}$ (b), and $\SI{0.2}{mm}$ (c). Tail to the left.}
   \label{f4}
\end{figure}

As demonstrated in Fig.~\ref{f2}, the beam tail has a large energy-spread~\cite{ding2016}. This means the peak current in the tail may be controlled with $R_{56}$ adjustments between the linac and wiggler in a dispersive section called dog-leg 2~\cite{emma2010a}. In Fig.~\ref{f4} we provide an example of wagging the beam tail in dog-leg 2. With the near-optimal $R_{56}$ of $\SI{-0.15}{mm}$ in Fig.~\ref{f4}(b), the modulation amplitude in the beam core is largest. The overcompressed beam tail of Fig.~\ref{f4}(a) and the undercompressed tail of Fig.~\ref{f4}(c) yield a smaller modulation amplitude in the beam core.

\begin{figure}[t]
   \centering
   \includegraphics{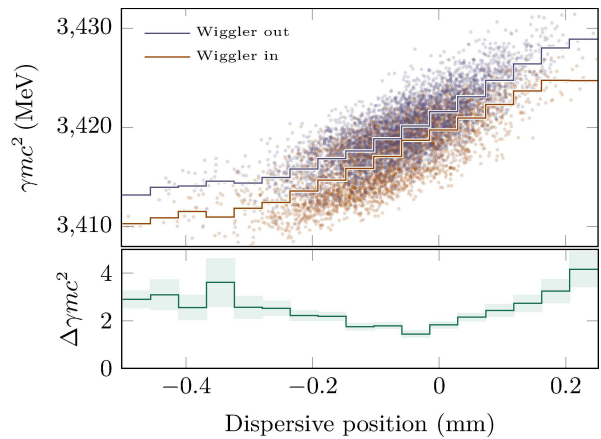}
\caption{(top) The average beam energy is plotted as a function of position in a dispersive region of the LTU when the wiggler is out (blue) and in (red). (bottom) The few MeV difference between the binned distributions is shown with $1-\sigma$ error bars.}
   \label{f5}
\end{figure}

A byproduct of the self-modulation process is a six-period, CEP-stable infrared light pulse at the resonant wavelength of the wiggler. The radiated pulse energy can be estimated by measuring the average energy loss of the electron-beam as it travels through the wiggler. Fig.~\ref{f5} shows the average bunch energy measured in the dump for 4000 consecutive shots with the wiggler set to $K=43.3$ (red) and $K=0$ (blue). The data are distributed along the horizontal axis according the beam position in a dispersive portion of the linac upstream from the wiggler. This helps distinguish between energy lost in the wiggler and shot-to-shot energy fluctuations that produce energy-correlated orbits. With the wiggler out, the average beam energy is larger by $\SI{2}{MeV}$ per electron, in rough agreement with the \SI{4.3}{MeV} energy loss from the idealized \textsc{osiris} simulation. We note that our setpoint optimized the stability of the beam-core energy modulation, but not the energy loss. We therefore see more energy loss at non-zero dispersive positions, where additional $R_{56}$ changes the current profile.

The $\SI{2}{MeV}$ energy loss and a charge of $\SI{140}{pC}$ imply an infrared pulse with $\SI{280}{\micro \joule}$, or roughly 4 gigawatts over six periods, is produced every shot. Note that, while the beam modulation exhibits a quasi-single-cycle temporal structure, our paraxial model predicts that the IR pulse is composed of 6 cycles with a uniform power profile in time.

$\SI{280}{\micro \joule}$ is comparable to dedicated infrared sources available to LCLS users~\cite{Minitti2015}. This pulse comes with sub-femtosecond timing-precision relative to the electron-beam at any linac repetition-rate. It is also possible to chirp the pulse with a wiggler taper, a feature that could be exploited for single-cycle infrared pulse production. 

In conclusion, we have demonstrated the generation of a phase-stable quasi-single-cycle infrared energy modulation of an electron bunch in a wiggler.
The modulation is induced by the interaction of the electrons with coherent radiation from the tail of the electron bunch. The quasi-single-cycle structure is largely due to strong diffraction along the wiggler, which lowers the field intensity experienced by the electrons far from the bunch tail.
The modulation is a few MeV in amplitude and stable in phase and period at the hundred attosecond level.

This method enables enhanced-SASE operation of LCLS for attosecond x-ray pulse production, a topic discussed elsewhere~\cite{duris2019}. 
The self-modulation process also results in the generation of a GW-scale CEP-stable infrared pulse that is timed to the electron bunch with sub-femtosecond stability.
Finally, this passive modulation method is applicable to the next generation of high-repetition rate, high average power X-ray free-electron lasers.

\begin{acknowledgments}
The authors thank C. Mayes, C. Emma, and G. Stupakov for fruitful discussions on modeling self-modulation. This work was supported by U.S. Department of Energy (DOE) contract No. DE-AC02-76SF00515, DOE-Basic Energy Sciences Field Work Proposal No. 100317, DOE Laboratory Directed Research and Development No. DE-AC02-76SF00515, and the Robert Siemann Graduate Fellowship.	The authors would like to acknowledge the OSIRIS Consortium, consisting of UCLA and IST (Lisbon, Portugal) for providing access to the OSIRIS 4.0 framework. Work supported by NSF ACI-1339893.
\end{acknowledgments}

%\bibliographystyle{apsrev4-2_10names}
%\bibliography{manuscript}

\begin{thebibliography}{38}%
\makeatletter
\providecommand \@ifxundefined [1]{%
 \@ifx{#1\undefined}
}%
\providecommand \@ifnum [1]{%
 \ifnum #1\expandafter \@firstoftwo
 \else \expandafter \@secondoftwo
 \fi
}%
\providecommand \@ifx [1]{%
 \ifx #1\expandafter \@firstoftwo
 \else \expandafter \@secondoftwo
 \fi
}%
\providecommand \natexlab [1]{#1}%
\providecommand \enquote  [1]{``#1''}%
\providecommand \bibnamefont  [1]{#1}%
\providecommand \bibfnamefont [1]{#1}%
\providecommand \citenamefont [1]{#1}%
\providecommand \href@noop [0]{\@secondoftwo}%
\providecommand \href [0]{\begingroup \@sanitize@url \@href}%
\providecommand \@href[1]{\@@startlink{#1}\@@href}%
\providecommand \@@href[1]{\endgroup#1\@@endlink}%
\providecommand \@sanitize@url [0]{\catcode `\\12\catcode `\$12\catcode
  `\&12\catcode `\#12\catcode `\^12\catcode `\_12\catcode `\%12\relax}%
\providecommand \@@startlink[1]{}%
\providecommand \@@endlink[0]{}%
\providecommand \url  [0]{\begingroup\@sanitize@url \@url }%
\providecommand \@url [1]{\endgroup\@href {#1}{\urlprefix }}%
\providecommand \urlprefix  [0]{URL }%
\providecommand \Eprint [0]{\href }%
\providecommand \doibase [0]{https://doi.org/}%
\providecommand \selectlanguage [0]{\@gobble}%
\providecommand \bibinfo  [0]{\@secondoftwo}%
\providecommand \bibfield  [0]{\@secondoftwo}%
\providecommand \translation [1]{[#1]}%
\providecommand \BibitemOpen [0]{}%
\providecommand \bibitemStop [0]{}%
\providecommand \bibitemNoStop [0]{.\EOS\space}%
\providecommand \EOS [0]{\spacefactor3000\relax}%
\providecommand \BibitemShut  [1]{\csname bibitem#1\endcsname}%
\let\auto@bib@innerbib\@empty
%</preamble>
\bibitem [{\citenamefont {Emma}\ \emph {et~al.}(2010)\citenamefont {Emma},
  \citenamefont {Akre}, \citenamefont {Arthur}, \citenamefont {Bionta},
  \citenamefont {Bostedt}, \citenamefont {Bozek}, \citenamefont {Brachmann},
  \citenamefont {Bucksbaum}, \citenamefont {Coffee}, \citenamefont {Decker},
  \citenamefont {Ding}, \citenamefont {Dowell}, \citenamefont {Edstrom},
  \citenamefont {Fisher}, \citenamefont {Frisch}, \citenamefont {Gilevich},
  \citenamefont {Hastings}, \citenamefont {Hays}, \citenamefont {Hering},
  \citenamefont {Huang}, \citenamefont {Iverson}, \citenamefont {Loos},
  \citenamefont {Messerschmidt}, \citenamefont {Miahnahri}, \citenamefont
  {Moeller}, \citenamefont {Nuhn}, \citenamefont {Pile}, \citenamefont
  {Ratner}, \citenamefont {Rzepiela}, \citenamefont {Schultz}, \citenamefont
  {Smith}, \citenamefont {Stefan}, \citenamefont {Tompkins}, \citenamefont
  {Turner}, \citenamefont {Welch}, \citenamefont {White}, \citenamefont {Wu},
  \citenamefont {Yocky},\ and\ \citenamefont {Galayda}}]{Emma2010}%
  \BibitemOpen
  \bibfield  {author} {\bibinfo {author} {\bibfnamefont {P.}~\bibnamefont
  {Emma}}, \bibinfo {author} {\bibfnamefont {R.}~\bibnamefont {Akre}}, \bibinfo
  {author} {\bibfnamefont {J.}~\bibnamefont {Arthur}}, \bibinfo {author}
  {\bibfnamefont {R.}~\bibnamefont {Bionta}}, \bibinfo {author} {\bibfnamefont
  {C.}~\bibnamefont {Bostedt}}, \bibinfo {author} {\bibfnamefont
  {J.}~\bibnamefont {Bozek}}, \bibinfo {author} {\bibfnamefont
  {A.}~\bibnamefont {Brachmann}}, \bibinfo {author} {\bibfnamefont
  {P.}~\bibnamefont {Bucksbaum}}, \bibinfo {author} {\bibfnamefont
  {R.}~\bibnamefont {Coffee}}, \bibinfo {author} {\bibfnamefont {F.-J.}\
  \bibnamefont {Decker}}, \bibnamefont {et~al.},\ }\href
  {https://doi.org/10.1038/nphoton.2010.176} {\bibfield  {journal} {\bibinfo
  {journal} {Nat. Photonics}\ }\textbf {\bibinfo {volume} {4}},\ \bibinfo
  {pages} {641} (\bibinfo {year} {2010})}\BibitemShut {NoStop}%
\bibitem [{\citenamefont {Ishikawa}\ \emph {et~al.}(2012)\citenamefont
  {Ishikawa}, \citenamefont {Aoyagi}, \citenamefont {Asaka}, \citenamefont
  {Asano}, \citenamefont {Azumi}, \citenamefont {Bizen}, \citenamefont {Ego},
  \citenamefont {Fukami}, \citenamefont {Fukui}, \citenamefont {Furukawa},
  \citenamefont {Goto}, \citenamefont {Hanaki}, \citenamefont {Hara},
  \citenamefont {Hasegawa}, \citenamefont {Hatsui}, \citenamefont {Higashiya},
  \citenamefont {Hirono}, \citenamefont {Hosoda}, \citenamefont {Ishii},
  \citenamefont {Inagaki}, \citenamefont {Inubushi}, \citenamefont {Itoga},
  \citenamefont {Joti}, \citenamefont {Kago}, \citenamefont {Kameshima},
  \citenamefont {Kimura}, \citenamefont {Kirihara}, \citenamefont {Kiyomichi},
  \citenamefont {Kobayashi}, \citenamefont {Kondo}, \citenamefont {Kudo},
  \citenamefont {Maesaka}, \citenamefont {Mar\'echal}, \citenamefont {Masuda},
  \citenamefont {Matsubara}, \citenamefont {Matsumoto}, \citenamefont
  {Matsushita}, \citenamefont {Matsui}, \citenamefont {Nagasono}, \citenamefont
  {Nariyama}, \citenamefont {Ohashi}, \citenamefont {Ohata}, \citenamefont
  {Ohshima}, \citenamefont {Ono}, \citenamefont {Otake}, \citenamefont {Saji},
  \citenamefont {Sakurai}, \citenamefont {Sato}, \citenamefont {Sawada},
  \citenamefont {Seike}, \citenamefont {Shirasawa}, \citenamefont {Sugimoto},
  \citenamefont {Suzuki}, \citenamefont {Takahashi}, \citenamefont {Takebe},
  \citenamefont {Takeshita}, \citenamefont {Tamasaku}, \citenamefont {Tanaka},
  \citenamefont {Tanaka}, \citenamefont {Tanaka}, \citenamefont {Togashi},
  \citenamefont {Togawa}, \citenamefont {Tokuhisa}, \citenamefont {Tomizawa},
  \citenamefont {Tono}, \citenamefont {Wu}, \citenamefont {Yabashi},
  \citenamefont {Yamaga}, \citenamefont {Yamashita}, \citenamefont {Yanagida},
  \citenamefont {Zhang}, \citenamefont {Shintake}, \citenamefont {Kitamura},\
  and\ \citenamefont {Kumagai}}]{Ishikawa2012}%
  \BibitemOpen
  \bibfield  {author} {\bibinfo {author} {\bibfnamefont {T.}~\bibnamefont
  {Ishikawa}}, \bibinfo {author} {\bibfnamefont {H.}~\bibnamefont {Aoyagi}},
  \bibinfo {author} {\bibfnamefont {T.}~\bibnamefont {Asaka}}, \bibinfo
  {author} {\bibfnamefont {Y.}~\bibnamefont {Asano}}, \bibinfo {author}
  {\bibfnamefont {N.}~\bibnamefont {Azumi}}, \bibinfo {author} {\bibfnamefont
  {T.}~\bibnamefont {Bizen}}, \bibinfo {author} {\bibfnamefont
  {H.}~\bibnamefont {Ego}}, \bibinfo {author} {\bibfnamefont {K.}~\bibnamefont
  {Fukami}}, \bibinfo {author} {\bibfnamefont {T.}~\bibnamefont {Fukui}},
  \bibinfo {author} {\bibfnamefont {Y.}~\bibnamefont {Furukawa}}, \bibnamefont
  {et~al.},\ }\href {https://doi.org/10.1038/nphoton.2012.141} {\bibfield
  {journal} {\bibinfo  {journal} {Nat. Photonics}\ }\textbf {\bibinfo {volume}
  {6}},\ \bibinfo {pages} {540} (\bibinfo {year} {2012})}\BibitemShut {NoStop}%
\bibitem [{\citenamefont {Ko}\ \emph {et~al.}(2017)\citenamefont {Ko},
  \citenamefont {Kang}, \citenamefont {Heo}, \citenamefont {Kim}, \citenamefont
  {Kim}, \citenamefont {Min}, \citenamefont {Yang}, \citenamefont {Baek},
  \citenamefont {Choi}, \citenamefont {Mun}, \citenamefont {Park},
  \citenamefont {Suh}, \citenamefont {Shin}, \citenamefont {Hu}, \citenamefont
  {Hong}, \citenamefont {Jung}, \citenamefont {Kim}, \citenamefont {Kim},
  \citenamefont {Na}, \citenamefont {Park}, \citenamefont {Park}, \citenamefont
  {Jung}, \citenamefont {Jeong}, \citenamefont {Lee}, \citenamefont {Lee},
  \citenamefont {Lee}, \citenamefont {Oh}, \citenamefont {Suh}, \citenamefont
  {Han}, \citenamefont {Kim}, \citenamefont {Jung}, \citenamefont {Kim},
  \citenamefont {Lee}, \citenamefont {Lee}, \citenamefont {Sung}, \citenamefont
  {Mok}, \citenamefont {Yang}, \citenamefont {Parc}, \citenamefont {Lee},
  \citenamefont {Lee}, \citenamefont {Shin}, \citenamefont {Kim}, \citenamefont
  {Kim}, \citenamefont {Lee}, \citenamefont {Park}, \citenamefont {Kim},
  \citenamefont {Park}, \citenamefont {Eom}, \citenamefont {Rah}, \citenamefont
  {Kim}, \citenamefont {Nam}, \citenamefont {Park}, \citenamefont {Park},
  \citenamefont {Kim}, \citenamefont {Kwon}, \citenamefont {An}, \citenamefont
  {Park}, \citenamefont {Kim}, \citenamefont {Hyun}, \citenamefont {Kim},
  \citenamefont {Kim}, \citenamefont {Yu}, \citenamefont {Kim}, \citenamefont
  {Kang}, \citenamefont {Kim}, \citenamefont {Kim}, \citenamefont {Lee},
  \citenamefont {Lee}, \citenamefont {Park}, \citenamefont {Koo}, \citenamefont
  {Kim},\ and\ \citenamefont {Lee}}]{Ko2017}%
  \BibitemOpen
  \bibfield  {author} {\bibinfo {author} {\bibfnamefont {I.}~\bibnamefont
  {Ko}}, \bibinfo {author} {\bibfnamefont {H.-S.}\ \bibnamefont {Kang}},
  \bibinfo {author} {\bibfnamefont {H.}~\bibnamefont {Heo}}, \bibinfo {author}
  {\bibfnamefont {C.}~\bibnamefont {Kim}}, \bibinfo {author} {\bibfnamefont
  {G.}~\bibnamefont {Kim}}, \bibinfo {author} {\bibfnamefont {C.-K.}\
  \bibnamefont {Min}}, \bibinfo {author} {\bibfnamefont {H.}~\bibnamefont
  {Yang}}, \bibinfo {author} {\bibfnamefont {S.}~\bibnamefont {Baek}}, \bibinfo
  {author} {\bibfnamefont {H.-J.}\ \bibnamefont {Choi}}, \bibinfo {author}
  {\bibfnamefont {G.}~\bibnamefont {Mun}}, \bibnamefont {et~al.},\ }\href
  {https://doi.org/10.3390/app7050479} {\bibfield  {journal} {\bibinfo
  {journal} {Appl. Sci.}\ }\textbf {\bibinfo {volume} {7}},\ \bibinfo {pages}
  {479} (\bibinfo {year} {2017})}\BibitemShut {NoStop}%
\bibitem [{\citenamefont {Milne}\ \emph {et~al.}(2017)\citenamefont {Milne},
  \citenamefont {Schietinger}, \citenamefont {Aiba}, \citenamefont {Alarcon},
  \citenamefont {Alex}, \citenamefont {Anghel}, \citenamefont {Arsov},
  \citenamefont {Beard}, \citenamefont {Beaud}, \citenamefont {Bettoni},
  \citenamefont {Bopp}, \citenamefont {Brands}, \citenamefont {Br\"onnimann},
  \citenamefont {Brunnenkant}, \citenamefont {Calvi}, \citenamefont {Citterio},
  \citenamefont {Craievich}, \citenamefont {Csatari~Divall}, \citenamefont
  {D\"allenbach}, \citenamefont {D'Amico}, \citenamefont {Dax}, \citenamefont
  {Deng}, \citenamefont {Dietrich}, \citenamefont {Dinapoli}, \citenamefont
  {Divall}, \citenamefont {Dordevic}, \citenamefont {Ebner}, \citenamefont
  {Erny}, \citenamefont {Fitze}, \citenamefont {Flechsig}, \citenamefont
  {Follath}, \citenamefont {Frei}, \citenamefont {G\"artner}, \citenamefont
  {Ganter}, \citenamefont {Garvey}, \citenamefont {Geng}, \citenamefont
  {Gorgisyan}, \citenamefont {Gough}, \citenamefont {Hauff}, \citenamefont
  {Hauri}, \citenamefont {Hiller}, \citenamefont {Humar}, \citenamefont
  {Hunziker}, \citenamefont {Ingold}, \citenamefont {Ischebeck}, \citenamefont
  {Janousch}, \citenamefont {Jurani\'c}, \citenamefont {Jurcevic},
  \citenamefont {Kaiser}, \citenamefont {Kalantari}, \citenamefont {Kalt},
  \citenamefont {Keil}, \citenamefont {Kittel}, \citenamefont {Knopp},
  \citenamefont {Koprek}, \citenamefont {Lemke}, \citenamefont {Lippuner},
  \citenamefont {Llorente~Sancho}, \citenamefont {L\"ohl}, \citenamefont
  {{Lopez-Cuenca}}, \citenamefont {M\"arki}, \citenamefont {Marcellini},
  \citenamefont {Marinkovic}, \citenamefont {Martiel}, \citenamefont {Menzel},
  \citenamefont {Mozzanica}, \citenamefont {Nass}, \citenamefont {Orlandi},
  \citenamefont {Ozkan~Loch}, \citenamefont {Panepucci}, \citenamefont
  {Paraliev}, \citenamefont {Patterson}, \citenamefont {Pedrini}, \citenamefont
  {Pedrozzi}, \citenamefont {Pollet}, \citenamefont {Pradervand}, \citenamefont
  {Prat}, \citenamefont {Radi}, \citenamefont {Raguin}, \citenamefont
  {Redford}, \citenamefont {Rehanek}, \citenamefont {R\'ehault}, \citenamefont
  {Reiche}, \citenamefont {Ringele}, \citenamefont {Rittmann}, \citenamefont
  {Rivkin}, \citenamefont {Romann}, \citenamefont {Ruat}, \citenamefont
  {Ruder}, \citenamefont {Sala}, \citenamefont {Schebacher}, \citenamefont
  {Schilcher}, \citenamefont {Schlott}, \citenamefont {Schmidt}, \citenamefont
  {Schmitt}, \citenamefont {Shi}, \citenamefont {Stadler}, \citenamefont
  {Stingelin}, \citenamefont {Sturzenegger}, \citenamefont {Szlachetko},
  \citenamefont {Thattil}, \citenamefont {Treyer}, \citenamefont {Trisorio},
  \citenamefont {Tron}, \citenamefont {Vetter}, \citenamefont {Vicario},
  \citenamefont {Voulot}, \citenamefont {Wang}, \citenamefont {Zamofing},
  \citenamefont {Zellweger}, \citenamefont {Zennaro}, \citenamefont {Zimoch},
  \citenamefont {Abela}, \citenamefont {Patthey},\ and\ \citenamefont
  {Braun}}]{Milne2017}%
  \BibitemOpen
  \bibfield  {author} {\bibinfo {author} {\bibfnamefont {C.}~\bibnamefont
  {Milne}}, \bibinfo {author} {\bibfnamefont {T.}~\bibnamefont {Schietinger}},
  \bibinfo {author} {\bibfnamefont {M.}~\bibnamefont {Aiba}}, \bibinfo {author}
  {\bibfnamefont {A.}~\bibnamefont {Alarcon}}, \bibinfo {author} {\bibfnamefont
  {J.}~\bibnamefont {Alex}}, \bibinfo {author} {\bibfnamefont {A.}~\bibnamefont
  {Anghel}}, \bibinfo {author} {\bibfnamefont {V.}~\bibnamefont {Arsov}},
  \bibinfo {author} {\bibfnamefont {C.}~\bibnamefont {Beard}}, \bibinfo
  {author} {\bibfnamefont {P.}~\bibnamefont {Beaud}}, \bibinfo {author}
  {\bibfnamefont {S.}~\bibnamefont {Bettoni}}, \bibnamefont {et~al.},\ }\href
  {https://doi.org/10.3390/app7070720} {\bibfield  {journal} {\bibinfo
  {journal} {Appl. Sci.}\ }\textbf {\bibinfo {volume} {7}},\ \bibinfo {pages}
  {720} (\bibinfo {year} {2017})}\BibitemShut {NoStop}%
\bibitem [{\citenamefont {Tschentscher}\ \emph {et~al.}(2017)\citenamefont
  {Tschentscher}, \citenamefont {Bressler}, \citenamefont {Gr\"unert},
  \citenamefont {Madsen}, \citenamefont {Mancuso}, \citenamefont {Meyer},
  \citenamefont {Scherz}, \citenamefont {Sinn},\ and\ \citenamefont
  {Zastrau}}]{Tschentscher2017}%
  \BibitemOpen
  \bibfield  {author} {\bibinfo {author} {\bibfnamefont {T.}~\bibnamefont
  {Tschentscher}}, \bibinfo {author} {\bibfnamefont {C.}~\bibnamefont
  {Bressler}}, \bibinfo {author} {\bibfnamefont {J.}~\bibnamefont {Gr\"unert}},
  \bibinfo {author} {\bibfnamefont {A.}~\bibnamefont {Madsen}}, \bibinfo
  {author} {\bibfnamefont {A.}~\bibnamefont {Mancuso}}, \bibinfo {author}
  {\bibfnamefont {M.}~\bibnamefont {Meyer}}, \bibinfo {author} {\bibfnamefont
  {A.}~\bibnamefont {Scherz}}, \bibinfo {author} {\bibfnamefont
  {H.}~\bibnamefont {Sinn}},\ \bibnamefont {and}\ \bibinfo {author}
  {\bibfnamefont {U.}~\bibnamefont {Zastrau}},\ }\href
  {https://doi.org/10.3390/app7060592} {\bibfield  {journal} {\bibinfo
  {journal} {Appl. Sci.}\ }\textbf {\bibinfo {volume} {7}},\ \bibinfo {pages}
  {592} (\bibinfo {year} {2017})}\BibitemShut {NoStop}%
\bibitem [{\citenamefont {Bostedt}\ \emph {et~al.}(2016)\citenamefont
  {Bostedt}, \citenamefont {Boutet}, \citenamefont {Fritz}, \citenamefont
  {Huang}, \citenamefont {Lee}, \citenamefont {Lemke}, \citenamefont {Robert},
  \citenamefont {Schlotter}, \citenamefont {Turner},\ and\ \citenamefont
  {Williams}}]{Bostedt2016}%
  \BibitemOpen
  \bibfield  {author} {\bibinfo {author} {\bibfnamefont {C.}~\bibnamefont
  {Bostedt}}, \bibinfo {author} {\bibfnamefont {S.}~\bibnamefont {Boutet}},
  \bibinfo {author} {\bibfnamefont {D.~M.}\ \bibnamefont {Fritz}}, \bibinfo
  {author} {\bibfnamefont {Z.}~\bibnamefont {Huang}}, \bibinfo {author}
  {\bibfnamefont {H.~J.}\ \bibnamefont {Lee}}, \bibinfo {author} {\bibfnamefont
  {H.~T.}\ \bibnamefont {Lemke}}, \bibinfo {author} {\bibfnamefont
  {A.}~\bibnamefont {Robert}}, \bibinfo {author} {\bibfnamefont {W.~F.}\
  \bibnamefont {Schlotter}}, \bibinfo {author} {\bibfnamefont {J.~J.}\
  \bibnamefont {Turner}},\ \bibnamefont {and}\ \bibinfo {author} {\bibfnamefont
  {G.~J.}\ \bibnamefont {Williams}},\ }\href
  {https://doi.org/10.1103/RevModPhys.88.015007} {\bibfield  {journal}
  {\bibinfo  {journal} {Rev. Mod. Phys.}\ }\textbf {\bibinfo {volume} {88}},\
  \bibinfo {pages} {015007} (\bibinfo {year} {2016})}\BibitemShut {NoStop}%
\bibitem [{\citenamefont {Kondratenko}\ and\ \citenamefont
  {Saldin}(1980)}]{Kondratenko1980}%
  \BibitemOpen
  \bibfield  {author} {\bibinfo {author} {\bibfnamefont {A.~M.}\ \bibnamefont
  {Kondratenko}}\ \bibnamefont {and}\ \bibinfo {author} {\bibfnamefont {E.~L.}\
  \bibnamefont {Saldin}},\ }\href@noop {} {\bibfield  {journal} {\bibinfo
  {journal} {Part Accel}\ }\textbf {\bibinfo {volume} {10}},\ \bibinfo {pages}
  {207} (\bibinfo {year} {1980})}\BibitemShut {NoStop}%
\bibitem [{\citenamefont {Bonifacio}\ \emph {et~al.}(1984)\citenamefont
  {Bonifacio}, \citenamefont {Pellegrini},\ and\ \citenamefont
  {Narducci}}]{Bonifacio1984}%
  \BibitemOpen
  \bibfield  {author} {\bibinfo {author} {\bibfnamefont {R.}~\bibnamefont
  {Bonifacio}}, \bibinfo {author} {\bibfnamefont {C.}~\bibnamefont
  {Pellegrini}},\ \bibnamefont {and}\ \bibinfo {author} {\bibfnamefont
  {L.}~\bibnamefont {Narducci}},\ }\href
  {https://doi.org/10.1016/0030-4018(84)90105-6} {\bibfield  {journal}
  {\bibinfo  {journal} {Opt. Commun.}\ }\textbf {\bibinfo {volume} {50}},\
  \bibinfo {pages} {373} (\bibinfo {year} {1984})}\BibitemShut {NoStop}%
\bibitem [{\citenamefont {Krausz}\ and\ \citenamefont
  {Ivanov}(2009)}]{Krausz2009}%
  \BibitemOpen
  \bibfield  {author} {\bibinfo {author} {\bibfnamefont {F.}~\bibnamefont
  {Krausz}}\ \bibnamefont {and}\ \bibinfo {author} {\bibfnamefont
  {M.}~\bibnamefont {Ivanov}},\ }\href
  {https://doi.org/10.1103/RevModPhys.81.163} {\bibfield  {journal} {\bibinfo
  {journal} {Rev. Mod. Phys.}\ }\textbf {\bibinfo {volume} {81}},\ \bibinfo
  {pages} {163} (\bibinfo {year} {2009})}\BibitemShut {NoStop}%
\bibitem [{\citenamefont {Bucksbaum}(2007)}]{Bucksbaum2007}%
  \BibitemOpen
  \bibfield  {author} {\bibinfo {author} {\bibfnamefont {P.~H.}\ \bibnamefont
  {Bucksbaum}},\ }\href {https://doi.org/10.1126/science.1142135} {\bibfield
  {journal} {\bibinfo  {journal} {Science}\ }\textbf {\bibinfo {volume}
  {317}},\ \bibinfo {pages} {766} (\bibinfo {year} {2007})}\BibitemShut
  {NoStop}%
\bibitem [{\citenamefont {Emma}\ \emph {et~al.}(2004)\citenamefont {Emma},
  \citenamefont {Bane}, \citenamefont {Cornacchia}, \citenamefont {Huang},
  \citenamefont {Schlarb}, \citenamefont {Stupakov},\ and\ \citenamefont
  {Walz}}]{Emma2004}%
  \BibitemOpen
  \bibfield  {author} {\bibinfo {author} {\bibfnamefont {P.}~\bibnamefont
  {Emma}}, \bibinfo {author} {\bibfnamefont {K.}~\bibnamefont {Bane}}, \bibinfo
  {author} {\bibfnamefont {M.}~\bibnamefont {Cornacchia}}, \bibinfo {author}
  {\bibfnamefont {Z.}~\bibnamefont {Huang}}, \bibinfo {author} {\bibfnamefont
  {H.}~\bibnamefont {Schlarb}}, \bibinfo {author} {\bibfnamefont
  {G.}~\bibnamefont {Stupakov}},\ \bibnamefont {and}\ \bibinfo {author}
  {\bibfnamefont {D.}~\bibnamefont {Walz}},\ }\href
  {https://doi.org/10.1103/PhysRevLett.92.074801} {\bibfield  {journal}
  {\bibinfo  {journal} {Phys. Rev. Lett.}\ }\textbf {\bibinfo {volume} {92}},\
  \bibinfo {pages} {074801} (\bibinfo {year} {2004})}\BibitemShut {NoStop}%
\bibitem [{\citenamefont {Huang}\ \emph {et~al.}(2017)\citenamefont {Huang},
  \citenamefont {Ding}, \citenamefont {Feng}, \citenamefont {Hemsing},
  \citenamefont {Huang}, \citenamefont {Krzywinski}, \citenamefont {Lutman},
  \citenamefont {Marinelli}, \citenamefont {Maxwell},\ and\ \citenamefont
  {Zhu}}]{Huang2017}%
  \BibitemOpen
  \bibfield  {author} {\bibinfo {author} {\bibfnamefont {S.}~\bibnamefont
  {Huang}}, \bibinfo {author} {\bibfnamefont {Y.}~\bibnamefont {Ding}},
  \bibinfo {author} {\bibfnamefont {Y.}~\bibnamefont {Feng}}, \bibinfo {author}
  {\bibfnamefont {E.}~\bibnamefont {Hemsing}}, \bibinfo {author} {\bibfnamefont
  {Z.}~\bibnamefont {Huang}}, \bibinfo {author} {\bibfnamefont
  {J.}~\bibnamefont {Krzywinski}}, \bibinfo {author} {\bibfnamefont {A.~A.}\
  \bibnamefont {Lutman}}, \bibinfo {author} {\bibfnamefont {A.}~\bibnamefont
  {Marinelli}}, \bibinfo {author} {\bibfnamefont {T.~J.}\ \bibnamefont
  {Maxwell}},\ \bibnamefont {and}\ \bibinfo {author} {\bibfnamefont
  {D.}~\bibnamefont {Zhu}},\ }\href
  {https://doi.org/10.1103/PhysRevLett.119.154801} {\bibfield  {journal}
  {\bibinfo  {journal} {Phys. Rev. Lett.}\ }\textbf {\bibinfo {volume} {119}},\
  \bibinfo {pages} {154801} (\bibinfo {year} {2017})}\BibitemShut {NoStop}%
\bibitem [{\citenamefont {Rosenzweig}\ \emph {et~al.}(2008)\citenamefont
  {Rosenzweig}, \citenamefont {Alesini}, \citenamefont {Andonian},
  \citenamefont {Boscolo}, \citenamefont {Dunning}, \citenamefont {Faillace},
  \citenamefont {Ferrario}, \citenamefont {Fukusawa}, \citenamefont
  {Giannessi}, \citenamefont {Hemsing}, \citenamefont {Marcus}, \citenamefont
  {Marinelli}, \citenamefont {Musumeci}, \citenamefont {O'Shea}, \citenamefont
  {Palumbo}, \citenamefont {Pellegrini}, \citenamefont {Petrillo},
  \citenamefont {Reiche}, \citenamefont {Ronsivalle}, \citenamefont {Spataro},\
  and\ \citenamefont {Vaccarezza}}]{ROSENZWEIG2008}%
  \BibitemOpen
  \bibfield  {author} {\bibinfo {author} {\bibfnamefont {J.}~\bibnamefont
  {Rosenzweig}}, \bibinfo {author} {\bibfnamefont {D.}~\bibnamefont {Alesini}},
  \bibinfo {author} {\bibfnamefont {G.}~\bibnamefont {Andonian}}, \bibinfo
  {author} {\bibfnamefont {M.}~\bibnamefont {Boscolo}}, \bibinfo {author}
  {\bibfnamefont {M.}~\bibnamefont {Dunning}}, \bibinfo {author} {\bibfnamefont
  {L.}~\bibnamefont {Faillace}}, \bibinfo {author} {\bibfnamefont
  {M.}~\bibnamefont {Ferrario}}, \bibinfo {author} {\bibfnamefont
  {A.}~\bibnamefont {Fukusawa}}, \bibinfo {author} {\bibfnamefont
  {L.}~\bibnamefont {Giannessi}}, \bibinfo {author} {\bibfnamefont
  {E.}~\bibnamefont {Hemsing}}, \bibnamefont {et~al.},\ }\href
  {https://doi.org/10.1016/j.nima.2008.04.083} {\bibfield  {journal} {\bibinfo
  {journal} {Nucl. Instrum. Methods Phys. Res. A}\ }\textbf {\bibinfo {volume}
  {593}},\ \bibinfo {pages} {39} (\bibinfo {year} {2008})}\BibitemShut
  {NoStop}%
\bibitem [{\citenamefont {Reiche}\ \emph {et~al.}(2008)\citenamefont {Reiche},
  \citenamefont {Musumeci}, \citenamefont {Pellegrini},\ and\ \citenamefont
  {Rosenzweig}}]{REICHE2008}%
  \BibitemOpen
  \bibfield  {author} {\bibinfo {author} {\bibfnamefont {S.}~\bibnamefont
  {Reiche}}, \bibinfo {author} {\bibfnamefont {P.}~\bibnamefont {Musumeci}},
  \bibinfo {author} {\bibfnamefont {C.}~\bibnamefont {Pellegrini}},\
  \bibnamefont {and}\ \bibinfo {author} {\bibfnamefont {J.}~\bibnamefont
  {Rosenzweig}},\ }\href {https://doi.org/10.1016/j.nima.2008.04.061}
  {\bibfield  {journal} {\bibinfo  {journal} {Nucl. Instrum. Methods Phys. Res.
  A}\ }\textbf {\bibinfo {volume} {593}},\ \bibinfo {pages} {45} (\bibinfo
  {year} {2008})}\BibitemShut {NoStop}%
\bibitem [{\citenamefont {Ding}\ \emph {et~al.}(2009)\citenamefont {Ding},
  \citenamefont {Brachmann}, \citenamefont {Decker}, \citenamefont {Dowell},
  \citenamefont {Emma}, \citenamefont {Frisch}, \citenamefont {Gilevich},
  \citenamefont {Hays}, \citenamefont {Hering}, \citenamefont {Huang},
  \citenamefont {Iverson}, \citenamefont {Loos}, \citenamefont {Miahnahri},
  \citenamefont {Nuhn}, \citenamefont {Ratner}, \citenamefont {Turner},
  \citenamefont {Welch}, \citenamefont {White},\ and\ \citenamefont
  {Wu}}]{Ding2009}%
  \BibitemOpen
  \bibfield  {author} {\bibinfo {author} {\bibfnamefont {Y.}~\bibnamefont
  {Ding}}, \bibinfo {author} {\bibfnamefont {A.}~\bibnamefont {Brachmann}},
  \bibinfo {author} {\bibfnamefont {F.-J.}\ \bibnamefont {Decker}}, \bibinfo
  {author} {\bibfnamefont {D.}~\bibnamefont {Dowell}}, \bibinfo {author}
  {\bibfnamefont {P.}~\bibnamefont {Emma}}, \bibinfo {author} {\bibfnamefont
  {J.}~\bibnamefont {Frisch}}, \bibinfo {author} {\bibfnamefont
  {S.}~\bibnamefont {Gilevich}}, \bibinfo {author} {\bibfnamefont
  {G.}~\bibnamefont {Hays}}, \bibinfo {author} {\bibfnamefont {P.}~\bibnamefont
  {Hering}}, \bibinfo {author} {\bibfnamefont {Z.}~\bibnamefont {Huang}},
  \bibnamefont {et~al.},\ }\href
  {https://doi.org/10.1103/PhysRevLett.102.254801} {\bibfield  {journal}
  {\bibinfo  {journal} {Phys. Rev. Lett.}\ }\textbf {\bibinfo {volume} {102}},\
  \bibinfo {pages} {254801} (\bibinfo {year} {2009})}\BibitemShut {NoStop}%
\bibitem [{\citenamefont {Prat}\ \emph {et~al.}(2015)\citenamefont {Prat},
  \citenamefont {L\"ohl},\ and\ \citenamefont {Reiche}}]{Prat2015}%
  \BibitemOpen
  \bibfield  {author} {\bibinfo {author} {\bibfnamefont {E.}~\bibnamefont
  {Prat}}, \bibinfo {author} {\bibfnamefont {F.}~\bibnamefont {L\"ohl}},\
  \bibnamefont {and}\ \bibinfo {author} {\bibfnamefont {S.}~\bibnamefont
  {Reiche}},\ }\href {https://doi.org/10.1103/PhysRevSTAB.18.100701} {\bibfield
   {journal} {\bibinfo  {journal} {Phys. Rev. ST Accel. Beams}\ }\textbf
  {\bibinfo {volume} {18}},\ \bibinfo {pages} {100701} (\bibinfo {year}
  {2015})}\BibitemShut {NoStop}%
\bibitem [{\citenamefont {Lutman}\ \emph {et~al.}(2016)\citenamefont {Lutman},
  \citenamefont {Maxwell}, \citenamefont {MacArthur}, \citenamefont {Guetg},
  \citenamefont {Berrah}, \citenamefont {Coffee}, \citenamefont {Ding},
  \citenamefont {Huang}, \citenamefont {Marinelli}, \citenamefont {Moeller},\
  and\ \citenamefont {Zemella}}]{Lutman2016}%
  \BibitemOpen
  \bibfield  {author} {\bibinfo {author} {\bibfnamefont {A.~A.}\ \bibnamefont
  {Lutman}}, \bibinfo {author} {\bibfnamefont {T.~J.}\ \bibnamefont {Maxwell}},
  \bibinfo {author} {\bibfnamefont {J.~P.}\ \bibnamefont {MacArthur}}, \bibinfo
  {author} {\bibfnamefont {M.~W.}\ \bibnamefont {Guetg}}, \bibinfo {author}
  {\bibfnamefont {N.}~\bibnamefont {Berrah}}, \bibinfo {author} {\bibfnamefont
  {R.~N.}\ \bibnamefont {Coffee}}, \bibinfo {author} {\bibfnamefont
  {Y.}~\bibnamefont {Ding}}, \bibinfo {author} {\bibfnamefont {Z.}~\bibnamefont
  {Huang}}, \bibinfo {author} {\bibfnamefont {A.}~\bibnamefont {Marinelli}},
  \bibinfo {author} {\bibfnamefont {S.}~\bibnamefont {Moeller}}, \bibnamefont
  {et~al.},\ }\href {https://doi.org/10.1038/nphoton.2016.201} {\bibfield
  {journal} {\bibinfo  {journal} {Nat. Photonics}\ }\textbf {\bibinfo {volume}
  {10}},\ \bibinfo {pages} {745} (\bibinfo {year} {2016})}\BibitemShut
  {NoStop}%
\bibitem [{\citenamefont {Saldin}\ \emph {et~al.}(2006)\citenamefont {Saldin},
  \citenamefont {Schneidmiller},\ and\ \citenamefont {Yurkov}}]{saldin2006}%
  \BibitemOpen
  \bibfield  {author} {\bibinfo {author} {\bibfnamefont {E.~L.}\ \bibnamefont
  {Saldin}}, \bibinfo {author} {\bibfnamefont {E.~A.}\ \bibnamefont
  {Schneidmiller}},\ \bibnamefont {and}\ \bibinfo {author} {\bibfnamefont
  {M.~V.}\ \bibnamefont {Yurkov}},\ }\href
  {https://doi.org/10.1103/PhysRevSTAB.9.050702} {\bibfield  {journal}
  {\bibinfo  {journal} {Phys. Rev. ST Accel. Beams}\ }\textbf {\bibinfo
  {volume} {9}},\ \bibinfo {pages} {050702} (\bibinfo {year}
  {2006})}\BibitemShut {NoStop}%
\bibitem [{\citenamefont {Zholents}(2005)}]{zholents2005}%
  \BibitemOpen
  \bibfield  {author} {\bibinfo {author} {\bibfnamefont {A.~A.}\ \bibnamefont
  {Zholents}},\ }\href {https://doi.org/10.1103/PhysRevSTAB.8.040701}
  {\bibfield  {journal} {\bibinfo  {journal} {Phys. Rev. ST Accel. Beams}\
  }\textbf {\bibinfo {volume} {8}},\ \bibinfo {pages} {040701} (\bibinfo {year}
  {2005})}\BibitemShut {NoStop}%
\bibitem [{\citenamefont {Tanaka}(2013)}]{Tanaka2013}%
  \BibitemOpen
  \bibfield  {author} {\bibinfo {author} {\bibfnamefont {T.}~\bibnamefont
  {Tanaka}},\ }\href {https://doi.org/10.1103/PhysRevLett.110.084801}
  {\bibfield  {journal} {\bibinfo  {journal} {Phys. Rev. Lett.}\ }\textbf
  {\bibinfo {volume} {110}},\ \bibinfo {pages} {084801} (\bibinfo {year}
  {2013})}\BibitemShut {NoStop}%
\bibitem [{\citenamefont {Shim}\ \emph {et~al.}(2018)\citenamefont {Shim},
  \citenamefont {Parc}, \citenamefont {Kumar}, \citenamefont {Ko},\ and\
  \citenamefont {Kim}}]{shim2018}%
  \BibitemOpen
  \bibfield  {author} {\bibinfo {author} {\bibfnamefont {C.~H.}\ \bibnamefont
  {Shim}}, \bibinfo {author} {\bibfnamefont {Y.~W.}\ \bibnamefont {Parc}},
  \bibinfo {author} {\bibfnamefont {S.}~\bibnamefont {Kumar}}, \bibinfo
  {author} {\bibfnamefont {I.~S.}\ \bibnamefont {Ko}},\ \bibnamefont {and}\
  \bibinfo {author} {\bibfnamefont {D.~E.}\ \bibnamefont {Kim}},\ }\href
  {https://doi.org/10.1038/s41598-018-25778-x} {\bibfield  {journal} {\bibinfo
  {journal} {Sci. Rep.}\ }\textbf {\bibinfo {volume} {8}},\ \bibinfo {pages}
  {7463} (\bibinfo {year} {2018})}\BibitemShut {NoStop}%
\bibitem [{\citenamefont {Hemsing}\ \emph {et~al.}(2014)\citenamefont
  {Hemsing}, \citenamefont {Stupakov}, \citenamefont {Xiang},\ and\
  \citenamefont {Zholents}}]{hemsing2014}%
  \BibitemOpen
  \bibfield  {author} {\bibinfo {author} {\bibfnamefont {E.}~\bibnamefont
  {Hemsing}}, \bibinfo {author} {\bibfnamefont {G.}~\bibnamefont {Stupakov}},
  \bibinfo {author} {\bibfnamefont {D.}~\bibnamefont {Xiang}},\ \bibnamefont
  {and}\ \bibinfo {author} {\bibfnamefont {A.}~\bibnamefont {Zholents}},\
  }\href {https://doi.org/10.1103/RevModPhys.86.897} {\bibfield  {journal}
  {\bibinfo  {journal} {Rev. Mod. Phys.}\ }\textbf {\bibinfo {volume} {86}},\
  \bibinfo {pages} {897} (\bibinfo {year} {2014})}\BibitemShut {NoStop}%
\bibitem [{\citenamefont {Fu}\ \emph {et~al.}(2018)\citenamefont {Fu},
  \citenamefont {Midorikawa},\ and\ \citenamefont {Takahashi}}]{fu2018}%
  \BibitemOpen
  \bibfield  {author} {\bibinfo {author} {\bibfnamefont {Y.}~\bibnamefont
  {Fu}}, \bibinfo {author} {\bibfnamefont {K.}~\bibnamefont {Midorikawa}},\
  \bibnamefont {and}\ \bibinfo {author} {\bibfnamefont {E.~J.}\ \bibnamefont
  {Takahashi}},\ }\href {https://doi.org/10.1038/s41598-018-25783-0} {\bibfield
   {journal} {\bibinfo  {journal} {Sci. Rep.}\ }\textbf {\bibinfo {volume}
  {8}},\ \bibinfo {pages} {7692} (\bibinfo {year} {2018})}\BibitemShut
  {NoStop}%
\bibitem [{\citenamefont {Saldin}\ \emph {et~al.}(1997)\citenamefont {Saldin},
  \citenamefont {Schneidmiller},\ and\ \citenamefont {Yurkov}}]{SALDIN1997}%
  \BibitemOpen
  \bibfield  {author} {\bibinfo {author} {\bibfnamefont {E.}~\bibnamefont
  {Saldin}}, \bibinfo {author} {\bibfnamefont {E.}~\bibnamefont
  {Schneidmiller}},\ \bibnamefont {and}\ \bibinfo {author} {\bibfnamefont
  {M.}~\bibnamefont {Yurkov}},\ }\href
  {https://doi.org/10.1016/S0168-9002(97)00822-X} {\bibfield  {journal}
  {\bibinfo  {journal} {Nucl. Instrum. Methods Phys. Res. A}\ }\textbf
  {\bibinfo {volume} {398}},\ \bibinfo {pages} {373} (\bibinfo {year}
  {1997})}\BibitemShut {NoStop}%
\bibitem [{\citenamefont {Saldin}\ \emph {et~al.}(1998)\citenamefont {Saldin},
  \citenamefont {Schneidmiller},\ and\ \citenamefont {Yurkov}}]{SALDIN1998}%
  \BibitemOpen
  \bibfield  {author} {\bibinfo {author} {\bibfnamefont {E.}~\bibnamefont
  {Saldin}}, \bibinfo {author} {\bibfnamefont {E.}~\bibnamefont
  {Schneidmiller}},\ \bibnamefont {and}\ \bibinfo {author} {\bibfnamefont
  {M.}~\bibnamefont {Yurkov}},\ }\href
  {https://doi.org/10.1016/S0168-9002(98)00623-8} {\bibfield  {journal}
  {\bibinfo  {journal} {Nucl. Instrum. Methods Phys. Res. A}\ }\textbf
  {\bibinfo {volume} {417}},\ \bibinfo {pages} {158} (\bibinfo {year}
  {1998})}\BibitemShut {NoStop}%
\bibitem [{\citenamefont {Wu}\ \emph {et~al.}(2003)\citenamefont {Wu},
  \citenamefont {Raubenheimer},\ and\ \citenamefont {Stupakov}}]{Wu2003}%
  \BibitemOpen
  \bibfield  {author} {\bibinfo {author} {\bibfnamefont {J.}~\bibnamefont
  {Wu}}, \bibinfo {author} {\bibfnamefont {T.~O.}\ \bibnamefont
  {Raubenheimer}},\ \bibnamefont {and}\ \bibinfo {author} {\bibfnamefont
  {G.~V.}\ \bibnamefont {Stupakov}},\ }\href
  {https://doi.org/10.1103/PhysRevSTAB.6.040701} {\bibfield  {journal}
  {\bibinfo  {journal} {Phys. Rev. ST Accel. Beams}\ }\textbf {\bibinfo
  {volume} {6}},\ \bibinfo {pages} {040701} (\bibinfo {year}
  {2003})}\BibitemShut {NoStop}%
\bibitem [{\citenamefont {Fonseca}\ \emph {et~al.}(2002)\citenamefont
  {Fonseca}, \citenamefont {Silva}, \citenamefont {Tsung}, \citenamefont
  {Decyk}, \citenamefont {Lu}, \citenamefont {Ren}, \citenamefont {Mori},
  \citenamefont {Deng}, \citenamefont {Lee}, \citenamefont {Katsouleas},\ and\
  \citenamefont {Adam}}]{Fonseca2002}%
  \BibitemOpen
  \bibfield  {author} {\bibinfo {author} {\bibfnamefont {R.~A.}\ \bibnamefont
  {Fonseca}}, \bibinfo {author} {\bibfnamefont {L.~O.}\ \bibnamefont {Silva}},
  \bibinfo {author} {\bibfnamefont {F.~S.}\ \bibnamefont {Tsung}}, \bibinfo
  {author} {\bibfnamefont {V.~K.}\ \bibnamefont {Decyk}}, \bibinfo {author}
  {\bibfnamefont {W.}~\bibnamefont {Lu}}, \bibinfo {author} {\bibfnamefont
  {C.}~\bibnamefont {Ren}}, \bibinfo {author} {\bibfnamefont {W.~B.}\
  \bibnamefont {Mori}}, \bibinfo {author} {\bibfnamefont {S.}~\bibnamefont
  {Deng}}, \bibinfo {author} {\bibfnamefont {S.}~\bibnamefont {Lee}}, \bibinfo
  {author} {\bibfnamefont {T.}~\bibnamefont {Katsouleas}}, \bibnamefont
  {et~al.},\ }in\ \href {https://doi.org/10.1007/3-540-47789-6_36} {\emph
  {\bibinfo {booktitle} {Computational {{Science}} \textemdash{} {{ICCS}}
  2002}}},\ Vol.\ \bibinfo {volume} {2331}\ (\bibinfo  {publisher} {{Springer
  Berlin Heidelberg}},\ \bibinfo {address} {Berlin, Heidelberg},\ \bibinfo
  {year} {2002})\ pp.\ \bibinfo {pages} {342--351}\BibitemShut {NoStop}%
\bibitem [{\citenamefont {Geloni}\ \emph {et~al.}(2007)\citenamefont {Geloni},
  \citenamefont {Saldin}, \citenamefont {Schneidmiller},\ and\ \citenamefont
  {Yurkov}}]{Geloni2007}%
  \BibitemOpen
  \bibfield  {author} {\bibinfo {author} {\bibfnamefont {G.}~\bibnamefont
  {Geloni}}, \bibinfo {author} {\bibfnamefont {E.}~\bibnamefont {Saldin}},
  \bibinfo {author} {\bibfnamefont {E.}~\bibnamefont {Schneidmiller}},\
  \bibnamefont {and}\ \bibinfo {author} {\bibfnamefont {M.}~\bibnamefont
  {Yurkov}},\ }\href {https://doi.org/10.1016/j.nima.2007.09.019} {\bibfield
  {journal} {\bibinfo  {journal} {Nucl. Instrum. Methods Phys. Res. A}\
  }\textbf {\bibinfo {volume} {583}},\ \bibinfo {pages} {228} (\bibinfo {year}
  {2007})}\BibitemShut {NoStop}%
\bibitem [{\citenamefont {Kim}\ \emph {et~al.}(2017)\citenamefont {Kim},
  \citenamefont {Huang},\ and\ \citenamefont {Lindberg}}]{kim2017}%
  \BibitemOpen
  \bibfield  {author} {\bibinfo {author} {\bibfnamefont {K.-J.}\ \bibnamefont
  {Kim}}, \bibinfo {author} {\bibfnamefont {Z.}~\bibnamefont {Huang}},\
  \bibnamefont {and}\ \bibinfo {author} {\bibfnamefont {R.}~\bibnamefont
  {Lindberg}},\ }\href {https://doi.org/10.1017/9781316677377} {\emph {\bibinfo
  {title} {Synchrotron {{Radiation}} and {{Free}}-{{Electron Lasers}}:
  {{Principles}} of {{Coherent X}}-{{Ray Generation}}}}}\ (\bibinfo
  {publisher} {{Cambridge University Press}},\ \bibinfo {address} {Cambridge},\
  \bibinfo {year} {2017})\BibitemShut {NoStop}%
\bibitem [{\citenamefont {Hofmann}(2004)}]{hofmann2004}%
  \BibitemOpen
  \bibfield  {author} {\bibinfo {author} {\bibfnamefont {A.}~\bibnamefont
  {Hofmann}},\ }\href {https://doi.org/10.1017/CBO9780511534973} {\emph
  {\bibinfo {title} {The {{Physics}} of {{Synchrotron Radiation}}}}}\ (\bibinfo
   {publisher} {{Cambridge University Press}},\ \bibinfo {address}
  {Cambridge},\ \bibinfo {year} {2004})\BibitemShut {NoStop}%
\bibitem [{\citenamefont {Wiedemann}(2007)}]{wiedemann2007}%
  \BibitemOpen
  \bibfield  {author} {\bibinfo {author} {\bibfnamefont {H.}~\bibnamefont
  {Wiedemann}},\ }\href {https://doi.org/10.1007/978-3-540-49045-6} {\emph
  {\bibinfo {title} {Particle {{Accelerator Physics}}}}},\ \bibinfo {edition}
  {3rd}\ ed.\ (\bibinfo  {publisher} {{Springer Berlin Heidelberg}},\ \bibinfo
  {address} {Berlin, Heidelberg},\ \bibinfo {year} {2007})\BibitemShut
  {NoStop}%
\bibitem [{\citenamefont {Geloni}\ \emph {et~al.}(2005)\citenamefont {Geloni},
  \citenamefont {Saldin}, \citenamefont {Schneidmiller},\ and\ \citenamefont
  {Yurkov}}]{Geloni2005}%
  \BibitemOpen
  \bibfield  {author} {\bibinfo {author} {\bibfnamefont {G.}~\bibnamefont
  {Geloni}}, \bibinfo {author} {\bibfnamefont {E.}~\bibnamefont {Saldin}},
  \bibinfo {author} {\bibfnamefont {E.}~\bibnamefont {Schneidmiller}},\
  \bibnamefont {and}\ \bibinfo {author} {\bibfnamefont {M.}~\bibnamefont
  {Yurkov}},\ }\href@noop {} {\bibfield  {journal} {\bibinfo  {journal}
  {arXiv:physics/0502120}\ } (\bibinfo {year} {2005})},\ \Eprint
  {https://arxiv.org/abs/physics/0502120} {arXiv:physics/0502120} \BibitemShut
  {NoStop}%
\bibitem [{\citenamefont {Behrens}\ \emph {et~al.}(2014)\citenamefont
  {Behrens}, \citenamefont {Decker}, \citenamefont {Ding}, \citenamefont
  {Dolgashev}, \citenamefont {Frisch}, \citenamefont {Huang}, \citenamefont
  {Krejcik}, \citenamefont {Loos}, \citenamefont {Lutman}, \citenamefont
  {Maxwell}, \citenamefont {Turner}, \citenamefont {Wang}, \citenamefont
  {Wang}, \citenamefont {Welch},\ and\ \citenamefont {Wu}}]{behrens2014}%
  \BibitemOpen
  \bibfield  {author} {\bibinfo {author} {\bibfnamefont {C.}~\bibnamefont
  {Behrens}}, \bibinfo {author} {\bibfnamefont {F.-J.}\ \bibnamefont {Decker}},
  \bibinfo {author} {\bibfnamefont {Y.}~\bibnamefont {Ding}}, \bibinfo {author}
  {\bibfnamefont {V.~A.}\ \bibnamefont {Dolgashev}}, \bibinfo {author}
  {\bibfnamefont {J.}~\bibnamefont {Frisch}}, \bibinfo {author} {\bibfnamefont
  {Z.}~\bibnamefont {Huang}}, \bibinfo {author} {\bibfnamefont
  {P.}~\bibnamefont {Krejcik}}, \bibinfo {author} {\bibfnamefont
  {H.}~\bibnamefont {Loos}}, \bibinfo {author} {\bibfnamefont {A.}~\bibnamefont
  {Lutman}}, \bibinfo {author} {\bibfnamefont {T.~J.}\ \bibnamefont {Maxwell}},
  \bibnamefont {et~al.},\ }\href {https://doi.org/10.1038/ncomms4762}
  {\bibfield  {journal} {\bibinfo  {journal} {Nat. Commun.}\ }\textbf {\bibinfo
  {volume} {5}},\ \bibinfo {pages} {3762} (\bibinfo {year} {2014})}\BibitemShut
  {NoStop}%
\bibitem [{\citenamefont {Ding}\ \emph {et~al.}(2016)\citenamefont {Ding},
  \citenamefont {Bane}, \citenamefont {Colocho}, \citenamefont {Decker},
  \citenamefont {Emma}, \citenamefont {Frisch}, \citenamefont {Guetg},
  \citenamefont {Huang}, \citenamefont {Iverson}, \citenamefont {Krzywinski},
  \citenamefont {Loos}, \citenamefont {Lutman}, \citenamefont {Maxwell},
  \citenamefont {Nuhn}, \citenamefont {Ratner}, \citenamefont {Turner},
  \citenamefont {Welch},\ and\ \citenamefont {Zhou}}]{ding2016}%
  \BibitemOpen
  \bibfield  {author} {\bibinfo {author} {\bibfnamefont {Y.}~\bibnamefont
  {Ding}}, \bibinfo {author} {\bibfnamefont {K.~L.~F.}\ \bibnamefont {Bane}},
  \bibinfo {author} {\bibfnamefont {W.}~\bibnamefont {Colocho}}, \bibinfo
  {author} {\bibfnamefont {F.-J.}\ \bibnamefont {Decker}}, \bibinfo {author}
  {\bibfnamefont {P.}~\bibnamefont {Emma}}, \bibinfo {author} {\bibfnamefont
  {J.}~\bibnamefont {Frisch}}, \bibinfo {author} {\bibfnamefont {M.~W.}\
  \bibnamefont {Guetg}}, \bibinfo {author} {\bibfnamefont {Z.}~\bibnamefont
  {Huang}}, \bibinfo {author} {\bibfnamefont {R.}~\bibnamefont {Iverson}},
  \bibinfo {author} {\bibfnamefont {J.}~\bibnamefont {Krzywinski}},
  \bibnamefont {et~al.},\ }\href
  {https://doi.org/10.1103/PhysRevAccelBeams.19.100703} {\bibfield  {journal}
  {\bibinfo  {journal} {Phys. Rev. Accel. Beams}\ }\textbf {\bibinfo {volume}
  {19}},\ \bibinfo {pages} {100703} (\bibinfo {year} {2016})}\BibitemShut
  {NoStop}%
\bibitem [{\citenamefont {Zhang}\ \emph {et~al.}(2019)\citenamefont {Zhang},
  \citenamefont {Duris}, \citenamefont {MacArthur}, \citenamefont {Huang},\
  and\ \citenamefont {Marinelli}}]{Zhang2019}%
  \BibitemOpen
  \bibfield  {author} {\bibinfo {author} {\bibfnamefont {Z.}~\bibnamefont
  {Zhang}}, \bibinfo {author} {\bibfnamefont {J.}~\bibnamefont {Duris}},
  \bibinfo {author} {\bibfnamefont {J.~P.}\ \bibnamefont {MacArthur}}, \bibinfo
  {author} {\bibfnamefont {Z.}~\bibnamefont {Huang}},\ \bibnamefont {and}\
  \bibinfo {author} {\bibfnamefont {A.}~\bibnamefont {Marinelli}},\ }\href
  {https://doi.org/10.1103/PhysRevAccelBeams.22.050701} {\bibfield  {journal}
  {\bibinfo  {journal} {Phys. Rev. Accel. Beams}\ }\textbf {\bibinfo {volume}
  {22}},\ \bibinfo {pages} {050701} (\bibinfo {year} {2019})}\BibitemShut
  {NoStop}%
\bibitem [{\citenamefont {Emma}(2010)}]{emma2010a}%
  \BibitemOpen
  \bibfield  {author} {\bibinfo {author} {\bibfnamefont {P.}~\bibnamefont
  {Emma}},\ }\href@noop {} {\bibfield  {journal} {\bibinfo  {journal}
  {Unpublished}\ } (\bibinfo {year} {2010})}\BibitemShut {NoStop}%
\bibitem [{\citenamefont {Minitti}\ \emph {et~al.}(2015)\citenamefont
  {Minitti}, \citenamefont {Robinson}, \citenamefont {Coffee}, \citenamefont
  {Edstrom}, \citenamefont {Gilevich}, \citenamefont {Glownia}, \citenamefont
  {Granados}, \citenamefont {Hering}, \citenamefont {Hoffmann}, \citenamefont
  {Miahnahri}, \citenamefont {Milathianaki}, \citenamefont {Polzin},
  \citenamefont {Ratner}, \citenamefont {Tavella}, \citenamefont {Vetter},
  \citenamefont {Welch}, \citenamefont {White},\ and\ \citenamefont
  {Fry}}]{Minitti2015}%
  \BibitemOpen
  \bibfield  {author} {\bibinfo {author} {\bibfnamefont {M.~P.}\ \bibnamefont
  {Minitti}}, \bibinfo {author} {\bibfnamefont {J.~S.}\ \bibnamefont
  {Robinson}}, \bibinfo {author} {\bibfnamefont {R.~N.}\ \bibnamefont
  {Coffee}}, \bibinfo {author} {\bibfnamefont {S.}~\bibnamefont {Edstrom}},
  \bibinfo {author} {\bibfnamefont {S.}~\bibnamefont {Gilevich}}, \bibinfo
  {author} {\bibfnamefont {J.~M.}\ \bibnamefont {Glownia}}, \bibinfo {author}
  {\bibfnamefont {E.}~\bibnamefont {Granados}}, \bibinfo {author}
  {\bibfnamefont {P.}~\bibnamefont {Hering}}, \bibinfo {author} {\bibfnamefont
  {M.~C.}\ \bibnamefont {Hoffmann}}, \bibinfo {author} {\bibfnamefont
  {A.}~\bibnamefont {Miahnahri}}, \bibnamefont {et~al.},\ }\href
  {https://doi.org/10.1107/S1600577515006244} {\bibfield  {journal} {\bibinfo
  {journal} {J. Synchrotron Rad.}\ }\textbf {\bibinfo {volume} {22}},\ \bibinfo
  {pages} {526} (\bibinfo {year} {2015})}\BibitemShut {NoStop}%
\bibitem [{\citenamefont {Duris}(2019)}]{duris2019}%
  \BibitemOpen
  \bibfield  {author} {\bibinfo {author} {\bibfnamefont {J.}~\bibnamefont
  {Duris}},\ }\href@noop {} {\bibfield  {journal} {\bibinfo  {journal} {Prep.}\
  } (\bibinfo {year} {2019})}\BibitemShut {NoStop}%
\end{thebibliography}


\newpage
\section{Supplemental Materials}
\subsection{Line-Charge Model}

In this section we arrive at Equation~\ref{emod} by working from the Li\'enard--Wiechert line-charge model~\cite{SALDIN1998}. This model describes the energy modulation of an electron-beam of line density $\lambda(s)$ propagating through an infinite planar wiggler. In this model the beam has no transverse extent, which produces an infinite space charge force that the authors subtract off~\cite{SALDIN1997} to recover tractability. What remains is both short and long-range curvature effects. Subsequent work~\cite{Wu2003} averaged the wake presented in~\cite{SALDIN1998} over a wiggler period for a large $K$ wiggler. 

Reference~\cite{Wu2003} describes the energy change after propagating a distance $z$ in an infinite wiggler,
\begin{equation}
    \eta(z,s) = \frac{e^2 }{\gamma m c^2} z W(s),\label{emodjuhao}
\end{equation}
with 
\begin{equation}
    W(s) = -k_u \int_{-\infty}^s ds' G(s-s') \frac{d \lambda(s')}{ds'},
\end{equation}
and 
\begin{equation}
    G(s) = \frac{2}{\pi}\int_0^\pi d\hat{z} \frac{\sin \Delta \cos \hat{z}+(1-\cos \Delta) \sin\hat{z}}{B(\Delta,\hat{z})},\label{beq}
\end{equation}
where $s$ is the longitudinal beam coordinate, $\Hat{z} = k_u z$, $\Delta = k_u (z-z_r)$ is the scaled distance between a test particle at $z$ and a source particle at the retarted position $z_r$. This scaled longitudinal distance can be determined by numerically solving the following transcendental equation for $\Delta$,
\begin{align}
    \zeta(\Delta,\Hat{z}) = \frac{\delta}{4} + \frac{1}{4\Delta} \{ &[2(1-\cos \Delta) - \Delta \sin \Delta]\nonumber\\ & \times (\cos \Delta \cos 2\Hat{z} + \sin \Delta \sin 2\Hat{z})\nonumber\\ &- 2(1-\cos \Delta) \}
\end{align}
where the beam coordinate $s$ has been scaled to $\zeta = s \gamma^2 k_u/K^2$. The $B(\Delta, \hat{z})$ in Equation~\ref{beq} can be calculated from $\Delta$,
\begin{align}
    B(\Delta, \Hat{z}) = &(1-\cos \Delta - \Delta \sin \Delta) \cos \hat{z} \nonumber \\ &+ 
    (\Delta \cos \Delta - \sin \Delta) \sin \hat{z}.
\end{align}
The behavior of $G(s)$ for $\zeta \gg 1$ was shown to contain harmonic content~\cite{Wu2003}
\begin{equation}
    G(\zeta) = -\frac{1}{2\zeta} + \frac{1}{2\zeta}\sum_{n=1}^{\infty} [\text{JJ}]_n^2 \cos(4(2n-1)\zeta),\label{gexp}
\end{equation}
where we have written harmonic coupling constant 
\begin{equation}
    [\text{JJ}]_n = J_{n-1}\left( \frac{2n-1}{2}\right) - J_{n}\left( \frac{2n-1}{2} \right) 
\end{equation}
in a way that suggests a direct connection to the paraxial approach in the main text. 

In order to recover Equation~\ref{emod} we calculate the energy modulation in response to an electron-beam of $N_e$ electrons at $s=0$, $\lambda(s) = N_e \delta(s)$. After integrating Equation~\ref{beq} by parts we find
\begin{equation}
    \eta(z,s) = \frac{r_e z k_u N_e}{\gamma} \frac{d G(s)}{ds},
\end{equation}
where the electron radius $r_e = e^2/mc^2$ in this unit system. Following the discussion in the main text about the duration of our current spike suppressing higher harmonic content, the lowest order term in Equation~\ref{gexp} can be differentiated to give
\begin{align}
    \eta(z,s) &\approx -\frac{2 r_e z k_u N_e}{\gamma s}  [\text{JJ}]_1^2 \sin (4 s \gamma^2 k_u / K^2)\\
    &= -\eta_0 \, k_u z \, \text{sinc} (k_1 s)\label{emodlong}
\end{align}
where terms of order $1/\zeta^2$ have been dropped in the first line, the resonant condition $k_1 \approx 4 k_u \gamma^2/K^2$ for large $K$ was used in the second line, and the amplitude $\eta_0$ from the main text was inserted. After identifying the ponderomotive phase with $k_1 s$, the only difference between Equation~\ref{emodlong} and Equation~\ref{emod} is in the term $(k_u z-\theta)$. This discrepancy is expected since the line-charge Li\'enard--Wiechert model assumes an infinitely long wiggler. We therefore conclude that the paraxial approach in the main text and the line-charge Li\'enard--Wiechert model agree in the core of the beam.

\subsection{Near-Axis Approximation}

\begin{figure}[htb]
   \centering
   \includegraphics{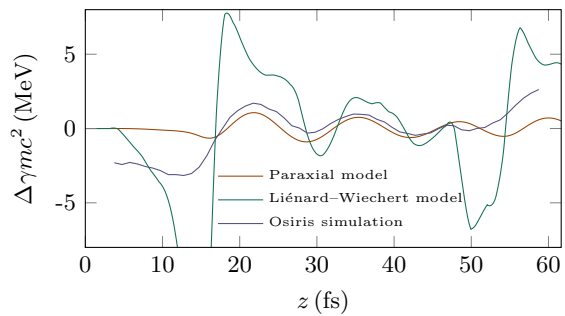}
\caption{A reproduction of Fig.~\ref{f2} with $\hat{\sigma}=3$. All other parameters are identical.}
   \label{bigsig}
\end{figure}

The importance of the scaled beam size, $\hat{\sigma}$, is demonstrated in Fig.~\ref{bigsig}. In this figure we reproduce the comparison between models and simulation from Fig.~\ref{f2} with a scaled beam size of $\hat{\sigma}=3.0$. All other parameters are the same.

The wide beam-size suppresses the modulation amplitude like $1/\hat{\sigma}^2$ near the source, a feature captured by Equation~\ref{emodfull}. As seen in Equation~\ref{emodfull}, the modulation phase also changes near the source for large $\hat{\sigma}$. This is a result of the Gouy phase-shift.

The Osiris simulation suggests that the paraxial model captures both of these effects correctly, while the line-charge model overestimates the modulation amplitude and yields an inconsistent modulation phase. 

In the remainder of this section we provide additional support for why the paraxial model produces the correct modulation amplitude and phase in the beam-core for a wide range of beam sizes.

\begin{figure}[htb]
   \centering
   \includegraphics{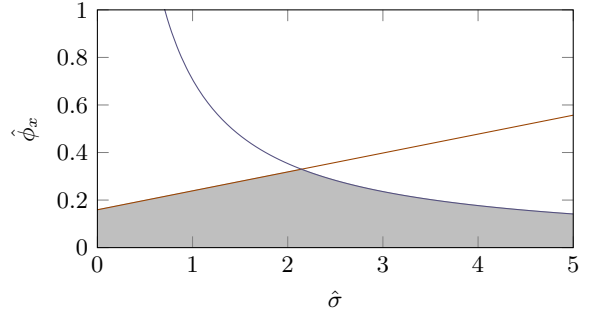}
\caption{Equation~\ref{rest1} (red), Equation~\ref{rest2} (blue), and the angular region described by these expressions (gray).}
   \label{restf}
\end{figure}

From the diagram in Fig.~\ref{f1} we can infer that only radiation forming at angles below $\sim (a_u + \sigma)/N\lambda_u$ is capable of interacting with the beam $N$ wavelengths from the source. If we are only interested in radiation in the beam-core, which is at least one wavelength from the tail, radiation is restricted to angles
\begin{equation}
    \phi_x < \frac{a_u + \sigma}{\lambda_u} = \frac{K}{\gamma} \left( \frac{1}{2\pi} + \frac{\sigma \gamma}{\lambda_u K} \right).
\end{equation}
The restriction for $\phi_y$ is stronger since the beam oscillates in the $xz$-plane, $\phi_y < \sigma/\lambda_u$. A natural scaling for angles in this problem is $K/\gamma$, the maximum angular deviation of an electron in a planar wiggler. The scaled angle $\hat{\phi}_x = \phi_x/(K/\gamma)$ is then restricted to
\begin{equation}
    \hat{\phi}_x < \frac{1}{2\pi} + \frac{\hat{\sigma}}{4\pi}.
    \label{rest1}
\end{equation}
This is a strong restriction when $\hat{\sigma} < 1$. When the scaled beam size is larger, as in Fig.~\ref{bigsig}, another restriction becomes important. The principle of superposition dictates that in the angular domain, emission at the fundamental wavelength $\lambda_1 = 2\pi / k_u$ from a Gaussian beam of width $\sigma_x = \sigma$ is suppressed like $\exp(-k_1^2 \sigma^2 \phi_x^2/2) = \exp(-2 \hat{\sigma}^2 \hat{\phi}_x^2)$. The restriction from this effect is therefore 
\begin{equation}
    \hat{\phi}_x < \frac{1}{\sqrt{2}\,\hat{\sigma}}.
    \label{rest2}
\end{equation}
The same inequality applies to $\phi_y$. In Fig.~\ref{restf}, Equations~\ref{rest1} and \ref{rest2} are shown bounding the region of angular space relevant to self-modulation. Emission at angles larger than $\hat{\phi}_x = 0.3298$ cannot participate in the modulation of the beam-core at any beam-size.

\begin{figure}[htb]
   \centering
   \includegraphics{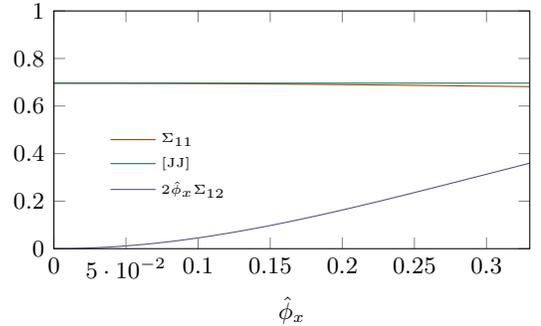}
\caption{A comparison of terms in Equation~\ref{fullfield} plotted as a function of the scaled angle $\hat{\phi}_x$. In this figure $\phi_y = 0$, and $1 \ll K \ll \gamma$.}
   \label{f9}
\end{figure}

The impact of these restrictions on the field relevant to self-modulation is best demonstrated by examining wiggler radiation in the far-field under the resonant approximation. When $K\gg 1$, Hofmann~\cite{hofmann2004} writes the radiation amplitude of the $m^{th}$ harmonic in the spectral domain as
\begin{align}
    \tilde{\mathbf{E}}_{\perp m} (\omega) \propto &\,\text{sinc}\left( \frac{\Delta \omega}{\omega_1(\phi)}\pi N_u\right)\nonumber \\
    &\times\left[ \left(2\hat{\phi}_x \Sigma_{m1} - \Sigma_{m2}\right)\hat{x} + 2\hat{\phi}_y \Sigma_{m1} \, \hat{y}\right],\label{fullfield}
\end{align}
where $\Delta \omega/\omega_1(\phi) = \omega (1+K^2/2+\phi^2\gamma^2)/(2ck_u \gamma^2)$ -1, 
\begin{align}
    \Sigma_{m1} &= \sum_{l=-\infty}^{\infty} J_l \left( m a_u \right) J_{m+2 l} \left( m b_u\right),\\
    \Sigma_{m2} &= \sum_{l=-\infty}^{\infty} \frac{2(m+2l)}{m b_u} J_l \left( m a_u \right) J_{m+2 l} \left( m b_u\right),
\end{align}
and 
\begin{align}
    a_u &= \frac{1}{2}\frac{1}{1+2\hat{\phi}^2}\\
    b_u &= 4\hat{\phi}_x\frac{1}{1+2\hat{\phi}^2}.
\end{align}
As discussed in the main text, higher harmonic content is suppressed by the finite extent of the tail current-spike. We therefore compare the terms $\Sigma_{11}$ and $2 \hat{\phi}_x \Sigma_{12}$ in Fig.~\ref{f9}. In this figure $\hat{\phi}_y = 0$ for simplicity. 
In the angular region relevant for self-modulation, $[\text{JJ}] =J_0(1/2) - J_1(1/2) \approx 0.696$ is an excellent approximation for $\Sigma_{11}$. We also observe that the term $2 \hat{\phi}_x \Sigma_{12}$ is significantly smaller than $[\text{JJ}]$ for most angles. The dominant contribution to the field capable of modulating the beam-core is therefore well described by the familiar $\text{sinc}$ in the frequency domain,
\begin{align}
    \tilde{\mathbf{E}}_{\perp m} (\omega) \approxprop &\,[\text{JJ}] \, \text{sinc} \left( \frac{\Delta \omega}{\omega_1(\phi)}\pi N_u\right) \hat{x}  \label{fullfieldapprox}.
\end{align}
Equation~\ref{fullfieldapprox} is the usual expression for the near-axis field from a beam in an undulator. It is therefore the solution to Equation~\ref{field} in the angular and frequency domain when $\mathbf{x}_j=0$ for frequencies near the fundamental. For example, see Equation 2.69 in Reference~\cite{kim2017}. We therefore find that solutions to Equation~\ref{field} will yield the field capable of participating in self-modulation.

\end{document}